\documentclass[runningheads]{llncs}
\usepackage[T1]{fontenc}
\usepackage{booktabs}
\usepackage{array}
\usepackage{amsmath}
\newcolumntype{C}[1]{>{\centering\arraybackslash}p{#1}}

\usepackage{graphicx}
\begin{document}
\title{A Human-Centered Approach to Ethical AI Education in Underresourced Secondary Schools}
\titlerunning{A Human-Centered Approach to Ethical AIED}
%
%
\author{Valentina Kuskova\inst{1,3}\orcidID{0000-0003-4716-2544} \and
Sonia Howell\inst{2,3}\orcidID{0000-0002-3368-9524} \and
Brianna Stines\inst{2,3} \and Brianna Conaghan\inst{2,3}}
\authorrunning{V. Kuskova et. al}
%
\institute{Lucy Family Institute for Data \& Society \and
Office of Digital Learning \and
University of Notre Dame, Notre Dame, IN, USA}
%
\maketitle              
\begin{abstract}
National and international policy efforts increasingly promote AI literacy in K–12 education, yet access to AI tools alone is insufficient to ensure equitable and responsible participation, particularly in under-resourced contexts. Without designs that integrate ethical reasoning, human support, and opportunities for judgment, AI initiatives risk reinforcing inequities and preparing students to use powerful technologies without critically evaluating their societal consequences.

To address this gap, we developed a human-centered, college credit–bearing course on Responsible and Ethical AI for students attending Title I and Title I–eligible high schools, implemented in partnership with the National Education Opportunity Network (NEON). The bichronous course integrates foundational AI concepts with ethical and socio-technical reasoning through asynchronous instruction, near-peer mentorship, and synchronous, discussion-based instruction. In its inaugural year, nearly 180 students from 12 U.S. schools completed the course, with a 97.8\% completion rate.

Using end-of-course survey data from students, co-teachers, and teaching fellows, this study examines academic agency, confidence with college-level expectations, critical engagement with responsible AI, and perceived post-secondary trajectory expansion. Results indicate that students were challenged to apply learning and reason about ethical tradeoffs, while educators reported high engagement, rigor, and meaningfulness relative to typical high school coursework. Overall, findings highlight the importance of human connections in advancing equitable Artificial Intelligence in Education and support ethical judgment as a core learning outcome alongside AI literacy.

\end{abstract}
\section{Introduction: AI Literacy, Equity, and the Limits of Access}
National and international policy efforts increasingly emphasize expanding artificial intelligence (AI) literacy in K–12 education, framing it as a necessary response to AI’s growing role in everyday life, work, and civic decision-making \cite{Karimov}, \cite{Milberg}, \cite{Malik}. In the United States, this emphasis has been formalized through executive action: an Executive Order dated April 23, 2025, in which President Trump \cite{Trump} directed a federal push to expand K–12 AI education through early exposure to AI concepts and strengthened educator training. Many initiatives prioritize student exposure to AI tools and applications \cite{Helikum}, \cite{Kwid}, reflecting the assumption that familiarity will prepare learners for future academic and workforce demands. At the same time, this policy momentum has raised critical questions about how AI education can support meaningful and responsible participation in an AI-mediated society, beyond technical awareness alone \cite{Ateşgöz}.

Within AIED research, a central concern is that access to AI tools and curricular content alone does not produce equitable outcomes, particularly in under-resourced educational contexts \cite{Schiff}. Schools serving students from low-income backgrounds often face constraints in instructional capacity, infrastructure, and sustained academic support \cite{Sucipto}, shaping how AI initiatives are implemented and experienced. This challenge is frequently described as the Artificial Intelligence in Education divide \cite{Ahmed}: a socio-technical gap that extends beyond disparities in technology or curriculum to include differences in instructional design, pedagogical support, and opportunities for guided reflection \cite{Memon}. From this perspective, equity in AI education depends not only on access, but on how learning experiences are structured and embedded within students’ educational trajectories \cite{Madaio}, including the social dimensions of learning such as mentorship, dialogue, and the cultivation of judgment in complex, real-world contexts.

Accordingly, scholars argue for AIED approaches that move beyond technical proficiency toward ethical judgment \cite{Biagini}. In under-resourced contexts, developing such judgment is closely tied to human support structures \cite{Kraft}, including instructors, mentors, and peers who create space for discussion and sense-making. Dialogic, socio-technical, and scaffolded learning design can therefore function as equity mechanisms in Artificial Intelligence in Education \cite{Bulathwela}.

Building on these insights, we developed a college credit-bearing course, \textit{Responsible \& Ethical AI} (\textit{R\&E AI course}), for students in Title I and Title I-eligible schools, offered in partnership with the National Education Opportunity Network (NEON) \cite{NEON}. The course operationalizes ethical judgment through human-centered learning in underresourced secondary school contexts. This paper reports findings from the first implementation, using end-of-course survey data from students, co-teachers (CTs), and teaching fellows (TFs) to examine academic agency, critical engagement with responsible AI, and preparation for postsecondary pathways. By centering ethical judgment, human connections, and learning design, this study demonstrates how equity-oriented AIED approaches can be operationalized and empirically examined beyond access-focused models.

\section{Related Work}
This study contributes to intersecting research streams in AIED, including AI literacy and critical AI education \cite{Yim}, equity and access in under-resourced contexts \cite{Samson}, and human-centered pedagogical approaches \cite{Ceresnova}. Rather than offering an exhaustive review, this section highlights key themes and gaps that motivate the design of the \textit{R\&E AI Course}.

\subsection{AI Literacy and Critical AI Education}
AI literacy \cite{Ng} has become a central focus of educational initiatives, commonly defined as understanding core AI concepts and systems \cite{Liu} and interacting effectively with AI-enabled tools \cite{Rzepka}. In K--12 contexts, AI literacy efforts typically emphasize exposure and foundational skills \cite{Chakraburty}, aiming to prepare students for AI-infused academic and workforce environments \cite{Bharathithasan}. These approaches have been important in establishing AI as a topic of general education \cite{Lee}.

More recently, scholars have called for a shift toward critical AI literacy \cite{Bridges}, which foregrounds ethical reasoning, power, and social impact \cite{Berson}. Critical AI education encourages learners to examine how AI systems are designed, deployed, and governed, and whose interests they serve \cite{Bao}, emphasizing AI’s role in reinforcing or challenging inequality \cite{Farahani}. Despite strong conceptual development, such approaches have been less frequently implemented in sustained K--12 settings \cite{Yue}, particularly outside well-resourced schools \cite{Hongli}.

Across both AI literacy and critical AI education, a persistent limitation is the emphasis on skills, exposure, or attitudinal change \cite{Yu}. Many programs assess conceptual understanding or tool familiarity \cite{Bergdahl} while giving limited attention to learners’ capacity for ethical judgment in realistic socio-technical contexts \cite{Yue}. Consequently, ethical reasoning and decision-making under uncertainty remain under-assessed and under-supported in AI education \cite{Lodhi}, with responsible action often assumed rather than empirically examined.

\subsection{Equity, Access, and the Artificial Intelligence in Education Divide}
Equity is a central concern in AIED research \cite{Holstein}, particularly as AI learning opportunities expand unevenly across educational systems. Under-resourced and low-resource school contexts—often serving students from low-income or historically marginalized communities—face persistent challenges in adopting and sustaining AI education, including limited infrastructure, instructional capacity, and access to specialized expertise \cite{Ahmed}, \cite{Bulathwela}.

Research emphasizes that inequity in AI education extends beyond access to tools or curricula to include differences in instructional support, contextualization, and integration into educational pathways \cite{Judijanto}, \cite{Nemani}. From this perspective, the Artificial Intelligence in Education divide reflects disparities in the quality and depth of engagement rather than the mere availability of AI-related content.

Accordingly, scholars caution that access-oriented approaches, particularly those focused on short-term exposure or tool provision, may exacerbate inequities \cite{Kim}. Without pedagogical scaffolding, mentorship, and opportunities for reflection \cite{Zou}, students in under-resourced contexts risk being positioned as passive users rather than critical participants \cite{Nasr}. Equity in AIED therefore requires attention to learning design and institutional support, not technological access alone.

\subsection{Human-Centered and Socio-Technical Approaches in AIED}
Human-centered approaches in AIED emphasize learners, educators, and social relationships as central to meaningful AI learning \cite{Giovanola}. Rather than treating AI education as purely technical, this work situates learning within socio-technical systems, recognizing that AI technologies are embedded in social practices, institutional structures, and value-laden decisions \cite{Herrmann}. Human-centered learning design therefore prioritizes interaction, dialogue, and responsiveness to learners’ contexts \cite{Troussas}.

Within this framework, dialogue and mentorship support ethical reasoning by enabling learners to articulate perspectives, confront ambiguity, and reason through tradeoffs \cite{Bustos}. These relational elements are especially important for ethical AI education, where outcomes center on judgment and responsibility rather than procedural mastery.

Despite growing recognition of these approaches, much AIED research continues to emphasize scalable, tool-driven, or asynchronous instruction \cite{Ferdousmou}, with limited attention to how human connections function as equity mechanisms in under-resourced settings. The course examined here addresses this gap by integrating ethical scaffolding, near-peer mentorship, and dialogic learning into a college credit–bearing AI course for Title I and Title I-eligible high schools. In doing so, it contributes empirical evidence on how ethical judgment and academic agency can be cultivated through human-centered learning design, extending current discussions of equity and ethics within AIED.

\section{Course Context and Partnership Model}

\textbf{Program Context and Learner Population.} The course was offered to students enrolled in twelve Title I and Title I-eligible high schools throughout the US, each serving a high proportion of students from low-income backgrounds \cite{Title I}. These schools are often situated in under-resourced educational environments characterized by constrained instructional capacity, uneven access to advanced coursework, and limited exposure to emerging fields such as artificial intelligence \cite{Garland}. School demographics are presented in Table 1, reflecting a geographically distributed and institutionally diverse set of under-resourced educational contexts.

\subsection{Partnership with the National Education Opportunity Network (NEON)}
The course was developed and implemented in partnership with the National Education Opportunity Network (NEON) \cite{NEON}, a national nonprofit organization focused on expanding educational opportunity and postsecondary access for students from underrepresented and low-income backgrounds (Title I and Title I–eligible schools). All courses are offered at no cost to students, supported through donations or school- and district-level subsidies.

This partnership was critical to enabling equitable and scalable AI education. Given that many high school educators have received little formal training in AI \cite{Lee22}, the NEON model allows university faculty to deliver AI instruction while preserving students’ existing classroom relationships and school contexts. The bichronous structure of a NEON course, combining synchronous and asynchronous components, enables learning to happen at scale while supporting consistent delivery across schools.

NEON also provided institutional infrastructure for recruitment, coordination with school-based educators, and alignment with college access initiatives, enabling integration within existing school environments while maintaining consistent instructional goals. Courses were delivered through the Canvas learning management system, with supplemental infrastructure (e.g., headphones) provided where needed. To ensure safe and equitable engagement, the course-offering institution\footnote{Will be de-anonymized after peer review.} provided access to a secure AI “sandbox” environment tailored to underresourced settings.

\begin{table}[htbp]
\centering
\caption{Student Demographics of Participating High Schools}
\label{tab:school_demographics}
\setlength{\tabcolsep}{3pt}
\renewcommand{\arraystretch}{1.15}
\begin{tabular}{C{2cm}C{1.4cm}C{1.2cm}C{1.4cm}C{1.2cm}C{1.2cm}C{1cm}C{1.2cm}}
\toprule
\textbf{State (School in Note)} &
\textbf{Total students} &
\textbf{Black / Afr.-Am.} &
\textbf{Hispanic / Latino} &
\textbf{White} &
\textbf{Asian} &
\textbf{Other} &
\textbf{\% Female} \\
\midrule
1. DC & 1,494 & 29\% & 68\% & -- & -- & $<3$\% & 46\% \\
2. NJ & 826 & 66\% & 31\% & -- & -- & $<3$\% & 44\% \\
3. NY & 383 & 24\% & 73\% & -- & -- & $<3$\% & 45\% \\
4. NY & 475 & -- & 100\% & -- & -- & -- & 45\% \\
5. IL & 721 & 95\% & 3\% & -- & -- & $<2$\% & 57\% \\
6. NJ & 865 & 36\% & 50\% & 10\% & -- & $<4$\% & 57\% \\
7. DC & 908 & 53\% & 45\% & -- & -- & $<2$\% & 44\% \\
8. NY & 311 & 38\% & 33\% & 5\% & 14\% & 10\% & 45\% \\
9. NY & 285 & 37\% & 52\% & 4\% & 3\% & $<4$\% & 67\% \\
10. CA & 2,176 & 3.5\% & 69\% & 12\% & 14\% & $<2$\% & 49\% \\
11. NY & 223 & 26\% & 47\% & 15\% & 8\% & $<4$\% & 56\% \\
12. FL & 1,106 & 1\% & 98\% & 1\% & -- & -- & 48\% \\
\bottomrule
\end{tabular}
{\footnotesize\textit{Note.}1. Columbia Heights Education Campus. 2. Central High School. 3. Claremont International High School. 4. International School for Liberal Arts 5. King College Prep 6. Newark School of Data Science \& Information Technology 7. Roosevelt High School 8. School for Classics High School 9. The Urban Assembly HS for Emergency Medicine 10. Van Nuys Senior High 11. Virtual Innovators Academy 12. Westland Hialeah Senior High School}
\end{table}

\section{Human-Centered Pedagogical Design}
The course operationalizes ethical AI education through a human-centered framework that prioritizes judgment, dialogue, and socio-technical reasoning. Ethics is not treated as an add-on to technical content; instead, it is integrated across learning objectives, instruction, and assessment. 

\subsection{Ethical Judgment as a Core Learning Objective}
A central design principle was prioritizing ethical judgment over technical proficiency. Students learned foundational AI concepts (e.g., machine learning, neural networks, data-driven decision-making, large language models) to evaluate real-world consequences rather than to master implementation. Ethical judgment was defined as reasoning about tradeoffs, competing values, and uncertainty. Students examined tensions such as efficiency versus fairness, innovation versus privacy, and automation versus human oversight, using ethical frameworks (e.g., deontological and consequentialist reasoning, fairness and responsibility principles) as decision-making tools rather than prescriptive answers.

This approach was reinforced through realistic scenarios drawn from domains such as hiring, healthcare, public services, social media, environmental impact of AI, and surveillance. Across cases, students were asked to articulate conditions for deployment, identify likely harms, and propose safeguards, emphasizing that responsible AI decisions are rarely reducible to a single correct solution.

\subsection{Instructional Model}
Instruction combined asynchronous content (approximately one hour of pre-recorded videos per week) with synchronous, dialogic learning through weekly Zoom sessions led by near-peer teaching fellows (TFs) and in-class lessons facilitated by school-based co-teachers (CTs). Live sessions emphasized interpretation, application, and reflection through guided discussion and small-group activities. This distributed instructional model integrated local contextual knowledge with postsecondary academic support: CTs oversaw classroom implementation and student engagement, while TFs - undergraduate students at the offering institution - led discussions, provided mentorship aligned with college-level expectations, and graded student work using standardized rubrics under the supervision of the course professor at the offering institution.

The curriculum progressed from foundational AI concepts and historical context to applied ethical analysis of contemporary AI deployments. Case-based and discussion-driven activities formed the core pedagogy, including ethical framework applications, policy debates, governance role-plays, and lifecycle mapping exercises that required students to articulate and defend positions. Assignments emphasized application and synthesis through written analyses, mini-projects, debates, and reflective essays addressing issues such as bias, privacy, explainability in high-stakes settings, automation, and generative AI. The course concluded with structured reflection in which students revisited their initial definitions of responsible and ethical AI and articulated how their thinking evolved. In its inaugural year, nearly 180 high school students completed the course, yielding a 97.8\% completion rate and indicating sustained engagement across the college-credit–bearing offering.

\section{Methods and Data}
This study uses a mixed-methods design drawing on end-of-course surveys from students, CTs, and TFs who participated in the inaugural implementation of the \textit{R\&E Course}. Surveys combined five-point Likert-scale items with open-ended questions, and were collected by NEON, anonymized, and analyzed at the aggregate level. Multiple stakeholder perspectives enabled triangulation of learning experiences, instructional dynamics, and perceived outcomes in under-resourced secondary school contexts.

Student surveys assessed academic agency, confidence with college-level expectations, engagement, critical thinking about responsible AI, and post-secondary aspirations, with open-ended reflections on course impact (122 responses; 68.2\% CTs). Co-teacher surveys captured perceptions of student engagement, rigor, meaningfulness, post-secondary readiness, and implementation challenges (8 responses; 75\% response rate). Teaching fellow surveys focused on mentorship experiences, professional growth, student engagement and challenge, and instructional support (12 responses; 86\% response rate).

\subsection{Analytic Approach}
The analytic strategy combined descriptive quantitative analysis with inductive qualitative thematic analysis, prioritizing transparency and triangulation over causal inference. Quantitative analyses summarized response distributions and patterns across stakeholder groups. Student survey items were aggregated into composite measures aligned with the study’s framework, including academic agency (confidence with college-level work, planning, and help-seeking) and judgment-oriented engagement (application of learning and critical engagement with multiple viewpoints). Composite construction emphasized conceptual coherence rather than model-based scaling.

Descriptive statistics (e.g., means and distributions) characterized student and educator responses, consistent with the study’s exploratory, program-evaluation focus. Open-ended responses were analyzed iteratively to identify recurring themes related to ethical reasoning, engagement, mentorship, academic confidence, perceived rigor, and equity-relevant supports. Themes were compared across stakeholder groups to assess convergence and divergence, strengthening interpretive validity through triangulation.

\section{Findings}
A summary of quantitative results for all stakeholder groups is presented in Table 2. Most measures present aggregated responses over several survey items. 
\subsection{Student Outcomes}
\textbf{Academic agency and confidence with college-level expectations.}
Students reported strong gains in academic agency and preparedness for college-level work ($\text{mean}=4.37$), including increased confidence handling college coursework, clearer understanding of academic expectations, greater willingness to seek help, and improved planning and time management. Open-ended responses attributed these gains to the rigor of assignments, exposure to college-style discussion, and feedback from TFs.
\begin{table}[htbp]
\centering
\caption{Descriptive Statistics of Survey Responses}
\label{tab:survey_descriptives}
\setlength{\tabcolsep}{6pt}
\renewcommand{\arraystretch}{1.15}
\begin{tabular}{lcccc}
\toprule
\textbf{Measure} & \textbf{Mean} & \textbf{St. Dev.} & \textbf{Min} & \textbf{Max} \\
\midrule
\multicolumn{5}{l}{\textbf{Student Responses}} \\
Academic agency & 4.37 & 0.74 & 1 & 5 \\
Course organization & 4.58 & 0.68 & 2 & 5 \\
Postsecondary trajectory expansion & 3.61 & 1.12 & 1 & 5 \\
Learning support & 4.59 & 0.68 & 2 & 5 \\
Comfortable learning environment & 4.79 & 0.50 & 3 & 5 \\
AI judgment-oriented engagement & 4.65 & 0.60 & 3 & 5 \\
\midrule
\multicolumn{5}{l}{\textbf{Co-Teacher Responses}} \\
Teaching enjoyment & 3.94 & 0.85 & 3 & 5 \\
Helping students prepare for college & 4.69 & 0.48 & 4 & 5 \\
Impact on students' postsecondary trajectory & 4.38 & 0.52 & 4 & 5 \\
Perceived student engagement &	4.25&	0.71&	3&	5\\
Perceived course meaningfulness to students	&4.88	&0.35	&4&	5\\
\midrule
\multicolumn{5}{l}{\textbf{Teaching Fellow Responses}} \\
Own valuable experience and career growth & 3.28 & 1.23 & 1 & 5 \\
Perceived student engagement & 3.92 & 1.00 & 1 & 5 \\
Perceived course meaningfulness to students & 3.92 & 1.08 & 1 & 5 \\
\bottomrule
\end{tabular}

\vspace{0.5em}
{\footnotesize\textit{Note.} All items were measured on five-point Likert scales, where $5$ is the "highest" or "completely agree."}
\end{table}

\textbf{Critical AI engagement and applied ethical reasoning.}
Course activities consistently challenged students to engage critically with responsible AI from multiple perspectives, yielding one of the highest survey ratings ($\text{mean}=4.65$). Through case analyses, debates, role-plays, and reflective writing, students evaluated AI applications not only for functionality or efficiency but also for fairness, privacy, accountability, and social impact. Peer dialogue and structured case discussions prompted students to reconsider initial assumptions, weigh risks and benefits, and justify decisions under uncertainty using course frameworks, reinforcing ethics as an applied, judgment-oriented reasoning process rather than abstract principle recall.

\textbf{Perceived post-secondary trajectory expansion.}
Students also reported shifts in how they viewed post-secondary options, with more moderate average ratings ($\text{mean}=3.61$, $\text{SD}=1.12$). Many indicated exploring colleges they had not previously considered or viewing more competitive institutions as attainable. Open-ended responses linked these shifts to increased confidence, exposure to college-level expectations, and interactions with TFs as near-peer role models. While self-reported, these findings suggest the course expanded students’ sense of academic possibility and preparedness.

\subsection{Educator and Teaching Fellow Perspectives}

\textbf{Perceived challenge, engagement, and meaningfulness.}
Co-teachers, more so than TFs, characterized the course as both meaningful ($\text{mean}=4.88$ vs. $3.92$) and engaging for students ($\text{mean}=4.25$ vs. $3.92$). Educators reported that the emphasis on discussion, written justification, and ethical analysis made the course more demanding than typical high school coursework, while also highly meaningful due to its relevance to contemporary societal issues and students’ lived experiences.

\textbf{Instructional rigor and role of mentorship.}
Relative to standard secondary curricula, educators perceived the course as offering higher cognitive demand and intellectual engagement, requiring students to grapple with ambiguity, defend positions, and revise reasoning. Teaching fellows observed that, with appropriate scaffolding, students were capable of sophisticated ethical reasoning, challenging assumptions about academic capacity in under-resourced contexts. Both CTs and TFs emphasized mentorship and instructional support as central to this rigor: synchronous sessions and near-peer mentoring built trust, encouraged participation, and modeled college-level academic discourse, while the combination of classroom support and external mentorship created an environment in which students felt comfortable asking questions and expressing uncertainty.

\subsection{Convergence and Divergence Across Stakeholders}
\textbf{Areas of convergence.}
Across students and CTs, there was strong alignment regarding the course’s rigor, engagement, and meaningfulness. Two groups on the high school level described the course as challenging yet accessible and emphasized the importance of discussion-based learning and mentorship. Student reports of applying learning and considering multiple viewpoints were corroborated by educator observations of sustained engagement.

\textbf{Areas of divergence and their significance.}
Differences emerged in emphasis across stakeholder groups: students most often highlighted personal growth, confidence, and a comfortable learning environment, while educators focused on instructional design and implementation challenges, and TFs emphasized relational dynamics. Moreover, TFs' ratings of course engagement and meaningfulness was somewhat lower relative to the other two groups. These differences reflect stakeholders’ distinct roles and perspectives and together provide a more comprehensive account of the course as both an educational intervention and a human-centered learning environment.

\section{Discussion}

This study examined a human-centered model of ethical AI education implemented in under-resourced secondary school contexts. The findings highlight how instructional designs that foreground ethical judgment and human support shape students’ engagement with AI beyond technical literacy, offering insights relevant to equity-focused Artificial Intelligence in Education.

\textbf{Ethical Judgment as a Core Outcome.}
A central contribution of this work is evidence that ethical judgment can be intentionally developed through AI education. Students were repeatedly asked to evaluate tradeoffs, consider multiple viewpoints, and justify decisions in realistic socio-technical contexts, indicating that judgment is a learnable practice rather than an incidental byproduct of AI exposure. These findings reinforce the limitations of AI literacy models that emphasize conceptual understanding or tool familiarity without addressing how learners reason about consequences, uncertainty, and responsibility. For AIED research, this underscores the importance of treating ethical judgment as an explicit and assessable learning outcome.

\textbf{Human-Centered Design and Equity.}
The findings also demonstrate that human connections, through mentorship, dialogue, and instructional support, advance equitable AI education in low-resource contexts. Near-peer mentorship and synchronous discussion supported student confidence, participation, and engagement with ethical complexity, functioning as equity mechanisms rather than auxiliary features. These results suggest that equity in AIED depends not only on access to tools or curricula, but on learning environments that provide relational support and guided sense-making.

\textbf{Implications for AIED Research and Practice.}
Together, these results suggest that equitable AI education requires integrating ethical judgment, human-centered pedagogy, and equity considerations from the outset. While scalable, asynchronous models can broaden access, they may be insufficient for outcomes that depend on dialogue, mentorship, and contextualized reasoning. This study demonstrates that hybrid approaches combining structured content with human support can advance policy goals for AI literacy while extending them beyond technical competence toward ethical responsibility and academic agency.

\section{Limitations and Future Work}

This study has limitations consistent with its exploratory focus on evaluating an implemented course rather than testing an intervention. Findings are based primarily on end-of-course, self-reported survey data, which capture participants’ perceptions and experiences but may be subject to response bias and do not constitute direct measures of learning or behavioral change. Although triangulation across students, co-teachers, and teaching fellows strengthens interpretive confidence, results should be understood as descriptive of how the course was experienced rather than as causal evidence of impact.

The study also lacks pre-course measures, limiting the ability to assess change over time. While participants frequently reported gains in confidence, ethical engagement, and academic agency, these perceptions cannot be compared against baseline levels. Future iterations of the course may incorporate pre- and post-course measures to better contextualize these findings.

Future work could also complement survey data with additional sources such as student artifacts, reflective writing, or discussion transcripts to provide deeper insight into how ethical judgment develops in practice. Longer-term follow-up could further illuminate how participation in human-centered AI education shapes students’ academic and post-secondary trajectories over time. Some of this work will be implemented in the next round of course to be offered in the Fall of 2026.

\section{Conclusion}
This paper examined a human-centered approach to ethical AI education implemented as a college-credit–bearing course for students in under-resourced secondary schools. Drawing on survey data from students, CTs, and TFs, the study contributes empirical evidence that ethical judgment, academic agency, and meaningful engagement with AI can be intentionally cultivated through instructional designs that prioritize dialogue, mentorship, and socio-technical reasoning. The findings demonstrate that students were not only exposed to AI concepts, but were also challenged to apply learning, consider multiple viewpoints, and reason through ethical tradeoffs in realistic contexts.

Collectively, the results support a reframing of AI education goals beyond AI literacy alone. While foundational knowledge of AI systems remains important, preparing students for responsible participation in an AI-mediated society requires attention to ethical judgment as a core learning outcome. This study shows that such judgment can be taught and practiced when learning environments are designed to support uncertainty, reflection, and interaction, particularly for students in low-resource contexts.

For Artificial Intelligence in Education research and practice, this work underscores the importance of human-centered, equity-focused design. As AI education initiatives continue to scale in response to policy imperatives, there is a risk that efficiency-driven, access-only models may overlook the relational and ethical dimensions necessary for equitable participation. By demonstrating how ethical judgment and human connections can be operationalized within a scalable educational partnership, this paper contributes to the literature discussing societal impact, ethical responsibility, and policy-aligned AI education. Ultimately, advancing equitable AIED requires not only expanding access to AI tools, but also designing learning experiences that center human judgment, responsibility, and opportunity.

\begin{credits}

\subsubsection{\discintname}
The authors have no competing interests to declare that are
relevant to the content of this article. 

\end{credits}
%
%
%
\bibliographystyle{splncs04}
%

\end{document}